\documentclass[10pt,conference]{IEEEtran} 

\usepackage{cite}
\usepackage{amsmath,amssymb,amsfonts}
\usepackage{algorithmic}
\usepackage{graphicx}
\usepackage{textcomp}
\usepackage{xcolor}
\usepackage{glossaries}
\usepackage{mathtools}

\begin{document}

\title{Deep Reinforcement Learning-Aided RAN Slicing Enforcement for B5G Latency Sensitive Services}

\author{\IEEEauthorblockN{Sergio Martiradonna\IEEEauthorrefmark{1}\IEEEauthorrefmark{4}, Andrea Abrardo\IEEEauthorrefmark{2}\IEEEauthorrefmark{4}, Marco Moretti\IEEEauthorrefmark{3}\IEEEauthorrefmark{4},Giuseppe Piro\IEEEauthorrefmark{1}\IEEEauthorrefmark{4}, and Gennaro Boggia\IEEEauthorrefmark{1}\IEEEauthorrefmark{4}}
\IEEEauthorblockA{\IEEEauthorrefmark{1}\textit{Dept. of Electrical and Information Engineering} - Politecnico di Bari, Bari, Italy\\
Email: \{name.surname\}@poliba.it}
\IEEEauthorblockA{\IEEEauthorrefmark{2}\textit{Dept. of Information Engineering and Applied Mathematics} - Università degli Studi di Siena, Siena, Italy\\
Email: abrardo@diism.unisi.it}
\IEEEauthorblockA{\IEEEauthorrefmark{3}\textit{Dept. of Information Engineering} - Università degli Studi di Pisa, Pisa, Italy\\
Email: marco.moretti@unipi.it}
\IEEEauthorblockA{\IEEEauthorrefmark{4}CNIT, Consorzio Nazionale Interuniversitario per le Telecomunicazioni}
}

\maketitle

\begin{abstract}
The combination of cloud computing capabilities at the network edge and artificial intelligence promise to turn future mobile networks into service- and radio-aware entities, able to address the requirements of upcoming latency-sensitive applications. In this context, a challenging research goal is to exploit edge intelligence to dynamically and optimally manage the Radio Access Network Slicing (that is a less mature and more complex technology than fifth-generation Network Slicing) and Radio Resource Management, which is a very complex task due to the mostly unpredictably nature of the wireless channel. This paper presents a novel architecture that leverages Deep Reinforcement Learning at the edge of the network in order to address Radio Access Network Slicing and Radio Resource Management optimization supporting latency-sensitive applications. The effectiveness of our proposal against baseline methodologies is investigated through computer simulation, by considering an autonomous-driving use-case.
\end{abstract}

\newacronym{umts}{UMTS}{Universal Mobile Telecommunications System} 
\newacronym{iot}{IoT}{Internet of Things} 
\newacronym{ioe}{IoE}{Internet of Everything} 
\newacronym{nb-iot}{NB-IoT}{NarrowBand IoT} 
\newacronym{m2m}{M2M}{Machine-to-Machine}
\newacronym{mtc}{MTC}{Machine-Type Communication}
\newacronym{emtc}{eMTC}{enhanced Machine Type Communication}
\newacronym{rat}{RAT}{Radio Access Technology}
\newacronym{ue}{UE}{User Equipment}
\newacronym{lte}{LTE}{Long Term Evolution}
\newacronym{ltea}{LTE-A}{LTE-Advanced}
\newacronym{rb}{RB} {Resource Block}
\newacronym{ul}{UL}{UpLink}
\newacronym{dl}{DL}{DownLink}
\newacronym{ofdm}{OFDM}{Orthogonal Frequency Division Multiplexing}
\newacronym{scfdm}{SC-FDM}{Single-Carrier Frequency Division Multiplexing}
\newacronym{ofdma}{OFDMA}{Orthogonal Frequency Division Multiple Access}
\newacronym{scfdma}{SC-FDMA}{Single-Carrier Frequency Division Multiple Access}
\newacronym{3gpp}{3GPP}{3rd Generation Partnership Project}
\newacronym{fdd}{FDD}{Frequency Division Duplex}
\newacronym{tdd}{TDD}{Time Division Duplex}
\newacronym{mme}{MME}{Mobility Management Entity}
\newacronym{eps}{EPS}{Evolved Packet System}
\newacronym{ciot}{CIoT}{Cellular Internet of Things}
\newacronym{srb}{SRB} {Signalling Radio Bearer}
\newacronym{drb}{DRB} {Data Radio Bearer}
\newacronym{rrc}{RRC} {Radio Resource Control}
\newacronym{nas}{NAS} {Non Access Stratum}
\newacronym{as}{AS} {Access Stratum}
\newacronym{po}{PO} {Paging Occasion}
\newacronym{pf}{PF} {Paging Frame}
\newacronym{ph}{PH} {Paging Hyperframe}
\newacronym{sgw}{S-GW} {Serving GateWay}
\newacronym{pgw}{P-GW} {Packet GateWay}
\newacronym{scef}{SCEF} {Service Capability Exposure Function}
\newacronym{enb}{eNB} {Evolved Node-B}
\newacronym{eutran}{E-UTRAN} {Evolved UMTS Terrestrial Radio Access Network}
\newacronym{utra}{UTRA} {UMTS Terrestrial Radio Access}
\newacronym{utran}{UTRAN} {UMTS Terrestrial Radio Access Network}
\newacronym{bpsk}{BPSK} {Binary Phase Shift Keying}
\newacronym{qpsk}{QPSK} {Quadrature Phase Shift Keying}
\newacronym{NPDCCH}{NPDCCH} {Narrowband Physical Downlink Control Channel}
\newacronym{pp}{PP} {PDCCH Period}
\newacronym{NPDSCH}{NPDSCH} {Narrowband Physical Downlink Shared Channel}
\newacronym{NPBCH}{NPBCH} {Narrowband Physical Broadcast Channel}
\newacronym{NPSS}{NPSS} {Narrowband Primary Synchronization Signal}
\newacronym{NSSS}{NSSS} {Narrowband Secondary Synchronization Signal}
\newacronym{NPUSCH}{NPUSCH} {Narrowband Physical Uplink Shared Channel}
\newacronym{NPRACH}{NPRACH} {Narrowband Physical Random Access Channel}
\newacronym{rar}{RAR} {Random Access Response}
\newacronym{sib}{SIB} {System Information Block}
\newacronym{mib}{MIB} {Master Information Block}
\newacronym{sibnb}{SIB-NB} {System Information Block - NarrowBand}
\newacronym{mibnb}{MIB-NB} {Master Information Block - NarrowBand}
\newacronym{tti}{TTI} {Transmission Time Interval}
\newacronym{harq}{HARQ} {Hybrid Automatic Repeat reQuest}
\newacronym{arq}{ARQ} {Automatic Repeat reQuest}
\newacronym{l2s}{L2S} {Link-To-System}
\newacronym{mrc}{MRC} {Maximum Ratio Combining}
\newacronym{tbs}{TBS} {Transport Block Size}
\newacronym{cp}{CP} {Cyclic Prefix}
\newacronym{pcid}{PCID} {Physical Cell ID}
\newacronym{cfo}{CFO} {Carrier Frequency Offset}
\newacronym{plmn}{PLMN} {Public Land Mobile Network}
\newacronym{drx}{DRX} {Discontinuous RX}
\newacronym{edrx}{eDRX} {enhanced Discontinuous RX}
\newacronym{imsi}{IMSI} {International Mobile Subscriber Identity}
\newacronym{psm}{PSM} {Power Saving Mode}
\newacronym{tau}{TAU} {Tracking Area Update}
\newacronym{rau}{RAU} {Routing Area Update}
\newacronym{sfn}{SFN} {System Frame Number}
\newacronym{hsfn}{H-SFN} {Hyper-System Frame Number}
\newacronym{rlf}{RLF} {Radio Link Failure}
\newacronym{mcl}{MCL} {Maximum Coupling Loss}
\newacronym{pdcp} {PDCP} {Packet Data Convergence Protocol}
\newacronym{amc} {AMC} {Adaptive Modulation and Coding}
\newacronym{qos} {QoS} {Quality of Service}
\newacronym{voip} {VoIP} {Voice over IP}
\newacronym{cdf} {CDF} {Cumulative Distribution Function}
\newacronym{ecdf} {ECDF} {Empirical Cumulative Distribution Function}
\newacronym{cbr} {CBR} {Constant BitRate}
\newacronym{udp} {UDP} {User Datagram Protocol}
\newacronym{mcs} {MCS} {Modulation and Coding Scheme}
\newacronym{mac} {MAC} {Media Access Control}
\newacronym{rlc} {RLC} {Radio Link Control}
\newacronym{ru} {RU} {Resource Unit}
\newacronym{dci} {DCI} {Downlink Control Indicator}
\newacronym{dvi} {DVI} {Data Volume Indicator}
\newacronym{sinr} {SINR} {Signal to Interference plus Noise Ratio}
\newacronym{dmrs} {DMRS} {DeModulation Reference Signal}
\newacronym{plr} {PLR} {Packet Loss Ratio}
\newacronym{cqi} {CQI} {Channel Quality Indicator}
\newacronym{crnti} {C-RNTI} {Cell Radio Network Temporary Identifier}
\newacronym{fifo} {FIFO} {First-In First-Out}
\newacronym{rr} {RR} {Round-Robin}
\newacronym{lpwa} {LPWA} {Low-Power Wide-Area}
\newacronym{bler} {BLER} {BLock Error Rate}
\newacronym{4g} {4G} {4th Generation}
\newacronym{5g} {5G} {5th Generation}
\newacronym{mimo} {MIMO} {Multiple-Input Multiple-Output}
\newacronym{kpi} {KPI} {Key Performance Indicator}
\newacronym{embb} {eMBB} {Enhanced Mobile Brodaband}
\newacronym{mmtc} {mMTC} {Massive Machine-Type Communications}
\newacronym{v2x} {V2X} {Vehicular to Everything}
\newacronym{urllc} {URLLC} {Ultra-Reliable Low-Latency Communications}
\newacronym{d2d} {D2D} {Device-to-Device}
\newacronym{tm} {TM} {Transmission Mode}

\newacronym{rlc-tm} {TM} {Transparent Mode}
\newacronym{rlc-um} {UM} {Uncknowledged Mode}
\newacronym{rlc-am} {AM} {Acknowledged Mode}

\newacronym{bwp} {BWP} {BandWidth Part}

\newacronym{pmi} {PMI} {Precoding Matrix Indicator}
\newacronym{ri} {RI} {Rank Indicator}
\newacronym{jsdm} {JSDM} {Joint Spatial Division and Multiplexing}
\newacronym{rzf} {RZF} {Regularized Zero-Forcing}
\newacronym{mbsfn} {MBSFN} {Multicast Broadcast Single Frequency Network}
\newacronym{srtapi} {SRTA-PI} {Separate Receive and Training Antennas with Polynomial Interpolation}
\newacronym{csi} {CSI} {Channel State Information}
\newacronym{rach}{RACH} {Random Access CHannel}
\newacronym{nr}{NR} {New Radio}
\newacronym{fec}{FEC} {Forward Error Correction}
\newacronym{mmse}{MMSE} {Minimum Mean Square Error}
\newacronym{miesm}{MIESM} {Mutual Information Effective SINR Mapping}
\newacronym{isd}{ISD} {Inter-Site Distance}
\newacronym{bms}{BMS} {Broadcast/Multicast Services}
\newacronym{uav}{UAV} {Unmanned Aerial Vehicle}
\newacronym{sdap}{SDAP}{Service Data Adaptation Protocol}

\newacronym{ai}{AI} {Artificial Intelligence}
\newacronym{b5g}{B5G} {Beyond 5G}

\newacronym{embb} {eMBB} {enhanced Mobile BroadBand}
\newacronym{mmtc} {mMTC} {massive Machine Type Communication}
\newacronym{urllc} {URLLC} {Ultra Reliable and Low Latency Communication}
\newacronym{ran} {RAN} {Radio Access Network}
\newacronym{cn} {CN} {core network}
\newacronym{cran} {C-RAN} {Cloud - Radio Access Network}
\newacronym{nr} {NR} {New Radio}
\newacronym{rrm} {RRM} {Radio Resource Management}
\newacronym{sla} {SLA} {Service Level Agreement}
\newacronym{v2x} {V2X} {Vehicle-to-Everything}
\newacronym{harq} {HARQ} {Hybrid Automatic Repeat reQuest}
\newacronym{mec} {MEC} {Multi-access Edge Computing}
\newacronym{ei} {EI} {Edge Intelligence}

\newacronym{5g} {5G} {Fifth Generation}
\newacronym{qos} {QoS} {Quality of Service}
\newacronym{noma} {NOMA} {Non Orthogonal Multiple Access}
\newacronym{tti} {TTI} {Transmission Time Interval}
\newacronym{nr} {NR} {New Radio}
\newacronym{mcs} {MCS} {Modulation and Coding Scheme}
\newacronym{diffserv}{DiffServ} {Differentiated Services}
\newacronym{qci} {QCI} {Quality of Service Class Identifiers}
\newacronym{rb}{RB} {Resource Block}
\newacronym{ott}{OTT} {Over-The-Top}
\newacronym{mvno}{MVNO} {Mobile Virtual Network Operator}
\newacronym{mce}{MCE} {Mobile Cloud Entity}
\newacronym{sdn}{SDN} {Software Defined Networking}
\newacronym{sme} {SME} {Slice Management Entity}
\newacronym{src} {SRC} {Slice Request Cache}
\newacronym{ip} {IP} {infrastructure provider}
\newacronym{mno} {MNO} {Mobile Network Operator}

\newacronym{api} {API} {Application Programming Interface}
\newacronym{paas} {PaaS} {Platform as a Service}
\newacronym{iaas} {IaaS} {Infrastructure as a Service}
\newacronym{rl} {RL} {Reinforcement Learning}
\newacronym{drl} {DRL} {Deep RL}
\newacronym{tnt} {TNT} {tenant}

\section{Introduction}
Cloud computing and caching capabilities at the edge, together with \gls{ai}, are 
two of the key enablers of next-generation mobile wireless communication systems, namely \gls{b5g} networks \cite{edge-intelligence}.  

One of the advantages of introducing intelligence at the edge is the capability of undertaking and executing dynamic \gls{ran} slicing.  Network Slicing is one of the major solutions for the management and integration of diverse applications with concurrent requirements \cite{foukas2017}. While slicing is already well-established in the Core segments of current 5G networks, \gls{ran} slicing is less mature and more challenging \cite{elayoubi2019}, mostly due to the nature of the wireless channel, which is unpredictably variable and prone to severe multi-user interference. 
For instance, the handling of ultra-reliable and low latency traffic is a specific example of the challenges met with \gls{ran} slicing \cite{martiradonna2019architecting}.  
In a conventional approach, these challenges can be met only at the cost of over-provisioning \gls{ran} resources, which are precious and scarce in general, and at the risk of disrupting other types of traffic. 
In \gls{b5g} networks, the \gls{ei} paradigm is de facto enforced to overcome the latency constraints, since not only data is to be processed in \gls{mec} servers, but also strategic scheduling decisions will be taken at the edge.

In particular, \gls{rl} and \gls{drl} appear to be particularly suited for addressing RAN slicing and \gls{rrm} optimization, which are problems where the optimum is very difficult to find, due to the non-convex nature of the resource allocation problems, and there is limited knowledge about the structure of the problems themselves \cite{sutton}. As a matter of fact, the roles of the different involved stakeholders should be maintained. 
In particular, the owner of the infrastructure should not be aware of third parties' most valuable information, while the latters will likely have an only partial understanding of the underlying RAN information, e.g., they are not authorized to precisely comprehend the procedures and algorithms implemented in the infrastructure and the internal functioning of the network apparatuses.
In this paper, we envision a virtualized control platform in which both the owners of the network infrastructure and third parties interact for enforcing Network Slices in the \gls{ran}.
Finally, focusing on the autonomous driving use case, we demonstrate the feasibility, the features, as well as the performance of the proposed architecture, thanks to computer simulations.

The remainder of this paper is organized as follows. Section \ref{RAN Slicing at the Edge} briefly introduces the RAN slicing problem and provides the rationale for using a reinforcement learning approach in the considered setting. Next, Section \ref{Reinforcement Learning for RAN slicing} describes the proposed reference architecture enabling dynamic RAN slicing at the edge by using a \gls{drl} algorithm.  Section \ref{A Case Study} presents a performance evaluation, discussing results and comparisons with alternative approaches. Finally, conclusions are drawn in Section \ref{Conclusions}.

\section{RAN Slicing and Reinforcement Learning}
\label{RAN Slicing at the Edge}

A prominent feature of \gls{ei} is \gls{ran} slicing: by exploiting the functionalities and capabilities of \gls{mec} resources, it is possible to partition the radio infrastructure to support orthogonal logical segments. Each network segment, or slice, provides a different service with its own \gls{sla} and the corresponding \gls{qos} requirements. Accordingly, the design of each slice is service-based, as it is steered by the requirements of a particular service \cite{foukas2017}. 
In the \gls{cn}, the operation of slicing allows the creation of segments with their own control and data plane functionalities, which are programmable and auto-configurable.
Typically, the slice \glspl{tnt}, i.e., the customers from vertical industries, have a vision of the underlying infrastructure as a virtualized entity of which they have, at least partially, control and which they can configure and operate independently \cite{Zhou2016}. Going along with this model, the \gls{ip} is the owner of the resources employed and the \gls{tnt} is allowed to use those resources, install its own applications, hold its own data, and enable its preferred security policies.
In the \gls{ran}, slicing is based on the virtualization of the radio resources and leverages software-defined networking (SDN) and network function virtualization (NFV) to create different end-to-end virtual networks over the same physical infrastructure. 
 Owing to the intrinsic shared and unpredictable nature of wireless resources, the integration in the \gls{ran} of the same attributes of slicing in the core network \gls{cn} is a complex task, and \gls{ran} slicing is a less mature and challenging practice, which involves several \gls{rrm} functionalities. e.g., spectrum planning, interference coordination, packet scheduling, and admission control \cite{Sallent2017}. 
The different solutions achieve different trade-offs between isolation and optimized resource utilization, between static and dynamic slice creation. 
For example, a static partitioning of the radio resources is not adequate to enforce the full vision of \gls{ran} slicing, as shown in \cite{Konstantinos2016}, which introduces the concept of a 5G network slice broker, designed to enable new players to dynamically request and lease resources from infrastructure providers via well-defined interfaces.

The advent of SDN and NFV also promotes AI-based resource allocation, management, and orchestration,  leading to full network automation, since large-scale data acquisition has become rather easier than before.
In order to devise an efficient slice enforcement strategy at the RAN level, two different prediction problems must be jointly addressed: (\emph{i}) prediction of the incoming agglomerated per-slice traffic; (\emph{ii})  prediction of the bandwidth efficiency at the radio link level. As for the first problem, solutions to anticipate future offered loads in mobile networks have been extensively studied in the last years, considering different approaches \cite{Nikravesh2016, Bega2020, Fu2016, Zhang2017, Li2014, Zhang2018, Gutterman2019}. For instance, in \cite{Bega2020}, a supervised deep learning approach for estimating the aggregate slice traffic at each data-center is presented and an ad-hoc cost function able to find a good trade-off between resource overprovisioning and service request denial is proposed. 
However, estimating the traffic represents only a partial step for the optimal slice resource allocation problem. Indeed, even in the presence of perfect traffic estimation, evaluating the optimal \gls{rrm} setting is a very difficult task owing to the random nature of the radio conditions. As a matter of fact, the problem of optimal \gls{rrm} is generally formulated as a non-convex optimization problem, whose solution requires the use of optimization tools with unmanageable computational complexity. 
Alternatively, ML in general, and RL in particular, has been recently investigated as a low-complexity  and  effective solution for \gls{rrm} in communication and computing systems \cite{Ye2019}, \cite{Faris2020}. 
In the RL framework, an RL agent can generate (near-) optimal control actions on the base of the immediate reward feedback from interactions with the environment. Together with simply optimizing the current reward in a greedy manner, the RL agent can take a long-term goal into account, which is essentially important to time-variant dynamic systems. 
Accordingly, RL appears to be particularly suited for \gls{rrm} problems when the following conditions hold \cite{sutton}:
\begin{enumerate}
    \item The optimum is unknown or very difficult to know, and only a reward associated with a given policy is available.
    \item The environment can be modeled as a Markov Decision Process (MDP), where, given a status that can be fully or partially observed by the agent, the reward depends deterministically or stochastically on the action taken by the agent. 
    \item The MDP model is not known or only partially known by the agent, e.g., the reward/loss function cannot be expressed in closed-form as a differentiable function of the allocation
decisions, i.e., the actions.
\end{enumerate}
As a matter of fact, all these conditions hold in the scenario considered in this paper, bringing out the capabilities of \gls{rl}- and \gls{drl}-based \gls{ei} for RAN slicing.
Moreover, differently from the already available studies \cite{Nikravesh2016, Bega2020,Ye2019, Faris2020}, we specifically consider the openness of the network to third parties, hence encouraging \glspl{tnt} to take (partial) control of the resources without deploying their own infrastructure. 

\section{The proposed Architecture for RAN Slicing with EI}
\label{Reinforcement Learning for RAN slicing}
\begin{figure}[!h]
	\centering
	\includegraphics[width=0.99\linewidth]{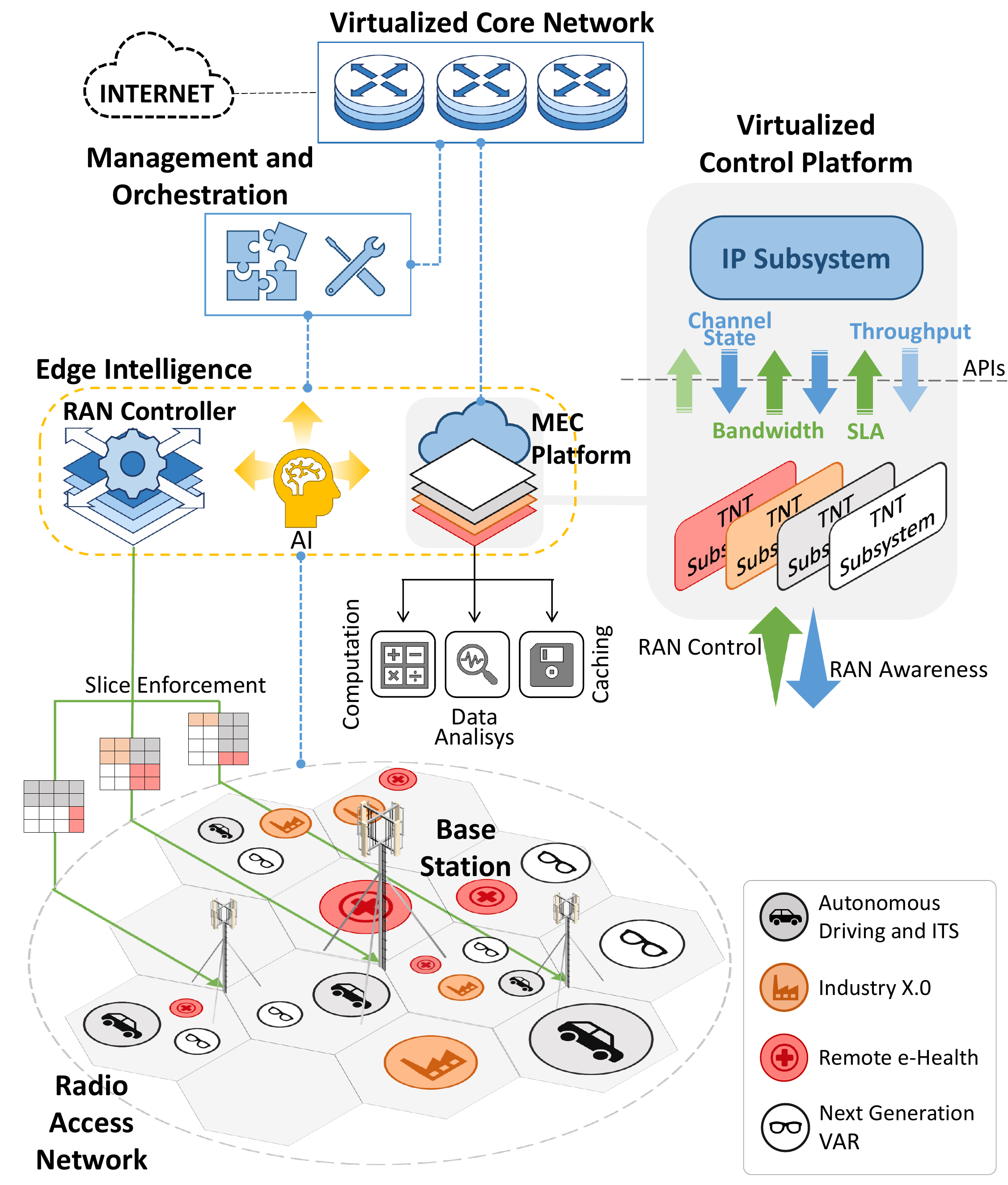}
	\caption{Reference architecture. As a reference, the picture shows the network slices of emerging latency-sensitive applications.}
	\label{fig:arch}
\end{figure}

We propose an architecture for RAN slicing by exploiting EI to serve mission critical applications.
To this end, in line with the current state of the art, we consider a scenario envisaging a single \gls{ip} that leases part of its network resources to create and manage specific slices for a set of independent \glspl{tnt}, or mobile network operators, to realize advanced network services \cite{Capone2019}. 

The \gls{ip} determines the amount of resources that can be used by the TNTs for each slice. The TNTs, in turn, should adapt in real-time their requests according to their own users' requirements, avoiding expenses due to the issue of resources overbuying. As a consequence, the generation of slice requests, i.e., when a TNT defines its needed slice configuration to the IP, and the slice dynamic enforcement, i.e., the adaptation of the slice allocation policy to the time-varying RAN environment, are the solutions of local optimization problems, which have to be solved in real-time and whose decisions have to be executed instantaneously to reduce any latency of the system. \cite{Doro2020}.

Fig. \ref{fig:arch} shows the reference architecture.
The general network scenario consists of a \gls{ran} and a Core Network. In the \gls{ran}, there are multiple BSs grouped in clusters that provide coverage to the service area. 
A  \gls{ran} controller for managing radio resources is located close to each of these clusters, together with a MEC platform, which provides the required computational and storage resources to implement functions requiring high real-time performance.

The controller dynamically performs slicing operations at the RAN layer, i.e., decides which slice creation requests can be admitted, computes a slicing policy to allocate the available resources to the admitted slices, and enforces the  slicing policy on the underlying physical RAN. 
 The  network slices are instantiated by the interactions of two different entities, the IP Subsystem and the TNT Subsystem. 
 The IP is the owner of this virtualized control platform and  dynamically leases computing and storage resources to TNTs to virtualize specific applications on the MEC. 
 Moreover, the IP Subsystem is in charge of creating different RAN slices leveraging on the \gls{ran} controller, according to the directives coming from the TNT subsystems. 
 To this aim, specific APIs are provided for the submission of slice requests. 
Hence, the main role of the IP is that of enforcing the slice request by allocating the required resources at the RAN layer through commands that are executed in the involved BSs, e.g., bandwidth reservation, number of BSs, interference coordination strategies among cells to fulfill inter-slice isolations, etc.  
The TNT subsystem essentially generates the slice requests by including general information, (e.g., type of services to be provided, the duration in time of the slice, ) as well as high-level control information for successfully addressing the requirements of the related slice. 
Requests are then sent to the IP Subsystem by means of the provided APIs. 
Hence, the IP must decide in advance the number of resources that will remain assigned to a slice until the next reallocation takes place. 

Since in this paper we focus on a mission-critical scenario, it is reasonable to assume that the TNT slice resource allocation requests must be always accepted by the IP, i.e., neither an admission control nor a resource allocation negotiation policy is enforced. Nevertheless, policies based on \emph{pay for what you get} mechanisms can be utilized by the IP to prevent from over-provisioning the TNT. As for the specific RAN slicing enforcement strategy, we then propose a dynamical RAN slicing at the Inter-cell Interference Coordination (ICIC) level, which is shown to provide high radio-electrical and traffic isolation with respect to the other slices \cite{Sallent2017}, that is one of the fundamental requirements for mission-critical services. To elaborate, each slice is assigned a given radio resource pool across a cluster of interfering cells in a given service area. The number of RBs is dynamically determined and requested by the TNT basing on the pieces of information it has access to. This scenario is in line with the general ambition of Network Slicing to open new business models for all the interested parties while maintaining the roles of the different involved stakeholders. As a consequence, the interaction of the participants must be regulated by APIs, where TNTs are not allowed to effectively comprehend the internal functioning of the devices provided by the IP and all the procedures implemented therein, as well as IPs should not be aware of TNTs' most valuable information.

The service level agreement between the latency-sensitive TNT and the IP provides for a unitary cost associated with each bandwidth resource and a maximum amount of bandwidth to be used in each cell. 
At the same time, based on some statistics the IP may choose to allocate the additional reserved bandwidth to outage-tolerant slices.

 \begin{figure}[!t]
	\centering
	\includegraphics[width=0.99\linewidth]{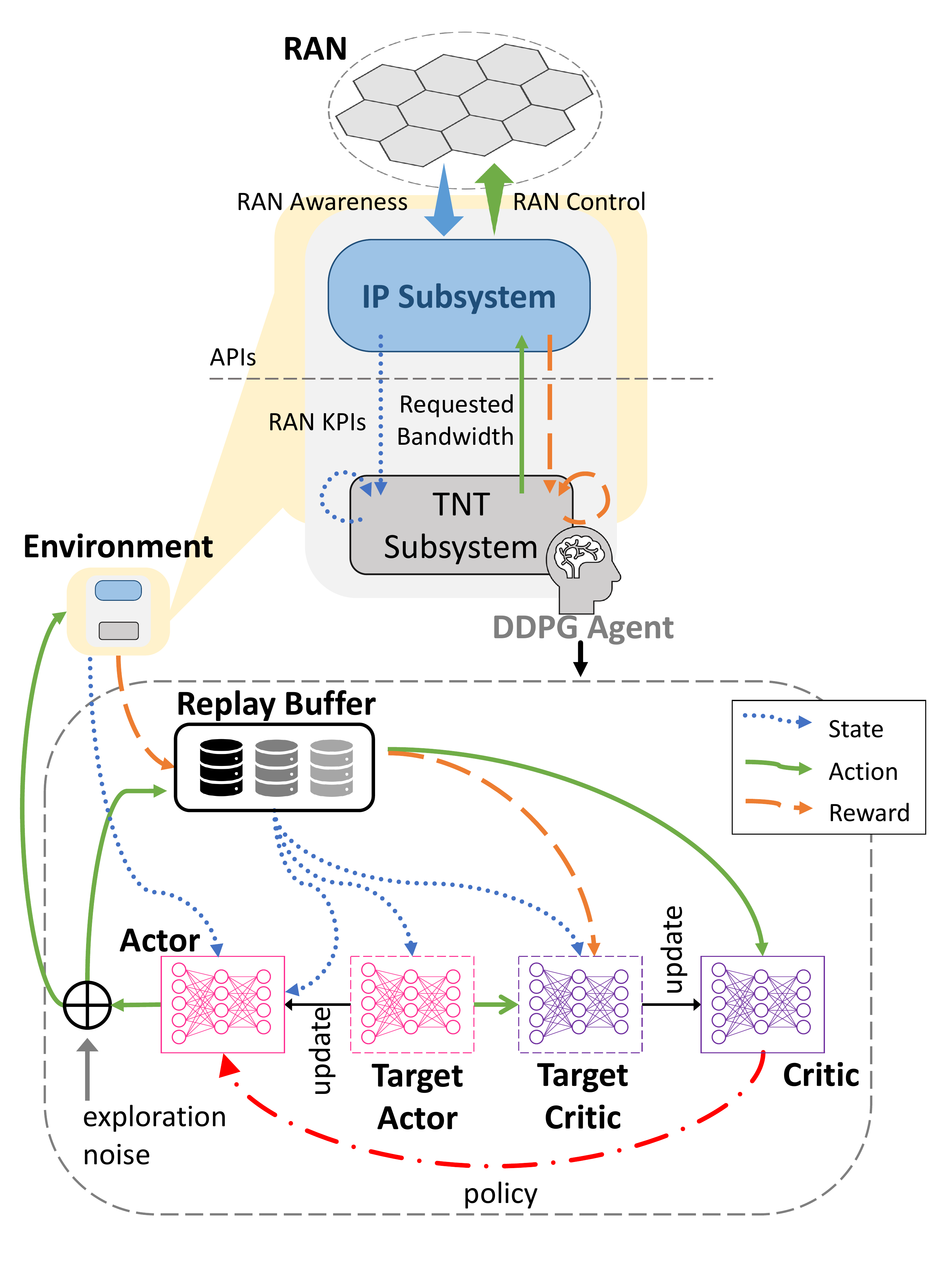}
	\caption{TNT agent and block diagram of the DDPG Algorithm.}
	\label{fig:ddpg}
\end{figure}

Starting from this architecture, it is evidently clear how \gls{rl} finds its applicability.
Indeed:
\begin{itemize}
    \item the scenario is a classical MDP where the \emph{environment} is the cellular system, the \emph{reward} is the efficiency of resource utilization subject to QoS constraints, which depends on the \emph{action}, i.e., the bandwidth allocated to mission critical slices, and on the radio conditions of the nodes, as well as the amount of incoming traffic, which altogether represent the \emph{state} of the environment;
    \item the optimal \gls{rrm} solution cannot be known because of the non-convex nature of the problem and for the fact that the TNTs have only partial knowledge of the underlying RAN information;
    \item the reward that is of interest for the TNT is often a QoS parameter, e.g., the latency and the packet loss ratio for mission-critical users, whose relationship with the allocation decision, e.g., the amount of allocated spectrum, is very hard to establish. 
\end{itemize}
In this setting, the role of the \gls{rl} agent of each TNT Subsystem is to reserve the minimum amount of bandwidth in each cell to satisfy its QoS requirements, so as to avoid resource overprovisioning.
The TNT places its bandwidth allocation requests expressed as a fraction of the maximum available bandwidth within a fixed allocation period (AP).
It is worth noting that the details of the radio interface, e.g., the adopted numerology, the scheduling policy, the packet fragmentation rules, and so on, are fully in charge of the IP and are not known by the TNT agents, which have only a limited knowledge of the radio link conditions of their users.

As illustrated in Fig. \ref{fig:ddpg}, a TNT agent is trained with the Deep Deterministic Policy Gradient (DDPG) algorithm, which is known to be suitable for dealing with continuous states and actions. Thanks to an actor-critic method, a DDPG agent concurrently learns a Q-function and an optimal policy that maximizes the long-term reward. DDPG exploits the Bellman equation to first learn the Q-function and then uses the Q-function to learn the optimal policy. Specifically, it is a model-free algorithm because the agent cannot predict the future states of the environment without taking the action. Besides, it is an off-policy method because the policy used to improve the Q-function approximation is different from the behavior policy, used to explore the environment. 

The \textit{action} is the amount of bandwidth requested to the IP every AP. The observations (or \textit{state}) that are available at the TNT are some Key Performance Indicators related to the \gls{ran} as well as traffic information.
Such observations can be either directly accessed by the TNT (e.g., the agglomerated slice traffic) or passed by the IP through the TNT-IP APIs. In addition, they are known at a given AP period and are used to evaluate the requested bandwidth for the next AP. 
The \textit{reward} should then take into account the amount of bandwidth the TNT saves with respect to the maximum bandwidth as well as some QoS indicators.

The amount of bandwidth is thus dynamically determined by the TNT basing on the available observations (state) with the goal of maximizing a discounted average future reward.

\section{Performance Evaluation}\label{A Case Study}
We evaluate the performance of a DRL TNT agent in a realistic environment where the MDP model is provided by a discrete event simulator of a cellular system developed in MATLAB. A single TNT subsystem is taken into account and it is assumed to provide an autonomous driving service. Without loss of generality, we focus on a single cell scenario, i.e., we do not consider the effect of inter-cell interference. On the other hand, the proposed framework leverages on the capability of the DRL agent to predict the mutual interactions of the involved nodes in determining the actual system performance and, accordingly, it is naturally suitable to encompass a multi-cell scenario, provided that the state variables include some interference related parameters, e.g., the mutual position of the nodes. 
We focus in the following on the downlink case. Similar considerations and results can be obtained for the uplink case.

Our scenario contains one single macro Base Station and a TNT subsystem, which provides autonomous driving services. In the considered setting, vehicles use their own sensors (e.g., HD camera, LiDAR), as well as sensor information from other vehicles, to perceive the environment and obtain a 3D model of the world around them.
The main QoS requirement of the slice is a maximum experienced packet delay of 5 ms, which is half the maximum value of latency envisioned for the High Definition Sensor Sharing, which is one of the main Autonomous Driving use cases \cite{5gaa}.
The packet length is assumed to be fixed and equal to 32 bytes (as per the Urban Macro–URLLC usage scenario \cite{itu}). 
The number of slice subscribers, i.e. the autonomous vehicles, is modeled according to real mobility traces \cite{dataset} and, on average, there are approximately 25 different vehicles in the cell.
Channel modeling considers path loss and lognormal shadowing. The data rate of each active link is derived based on the Shannon capacity formula.
The TNT is allocated a maximum bandwidth of 10 MHz, organized into slots of 1 ms, according to the 5G NR numerology with $\Delta f = 15$ KHz.
The MAC scheduling strategy enforced by the IP is the Throughput to Average scheduling, in order to guarantee a minimum level of service to every user, hence reaching a high fairness index.
The AP is 1 second, i.e., the DDPG agent performs its actions every second.

The \textit{action}, which is a continuous value between 0.1 and 0.9 (i.e., 10\% and 90\%), is the amount of bandwidth requested to the IP every AP. The \textit{state} is characterized by the maximum buffer occupancy experienced by the users of the slice, the total per-slice agglomerated traffic, and the list of the worst channel quality indicators (CQI) of the served users averaged over the AP.
The \textit{reward} $R$ is computed as:
\begin{equation*}
R= 
    \begin{dcases}
    1-b,& \text{if QoS requirement is satisfied}\\
    -1,              & \text{otherwise}
\end{dcases}
\end{equation*}
where $b$ is the ratio between the amount of bandwidth the TNT requests and the maximum bandwidth of 10 MHz. In other words, the less the bandwidth requested by the TNT, the higher the reward.

As customary in machine learning, the dataset of mobility traces considered during the training phase is different from the one considered during the testing phase. In this way, it is possible to assess the capability of the proposed DRL approach to generalize the proposed control strategy to every possible data traffic and radio channel conditions.  

 \begin{figure}[!h]
	\centering
	\includegraphics[width=0.85\linewidth]{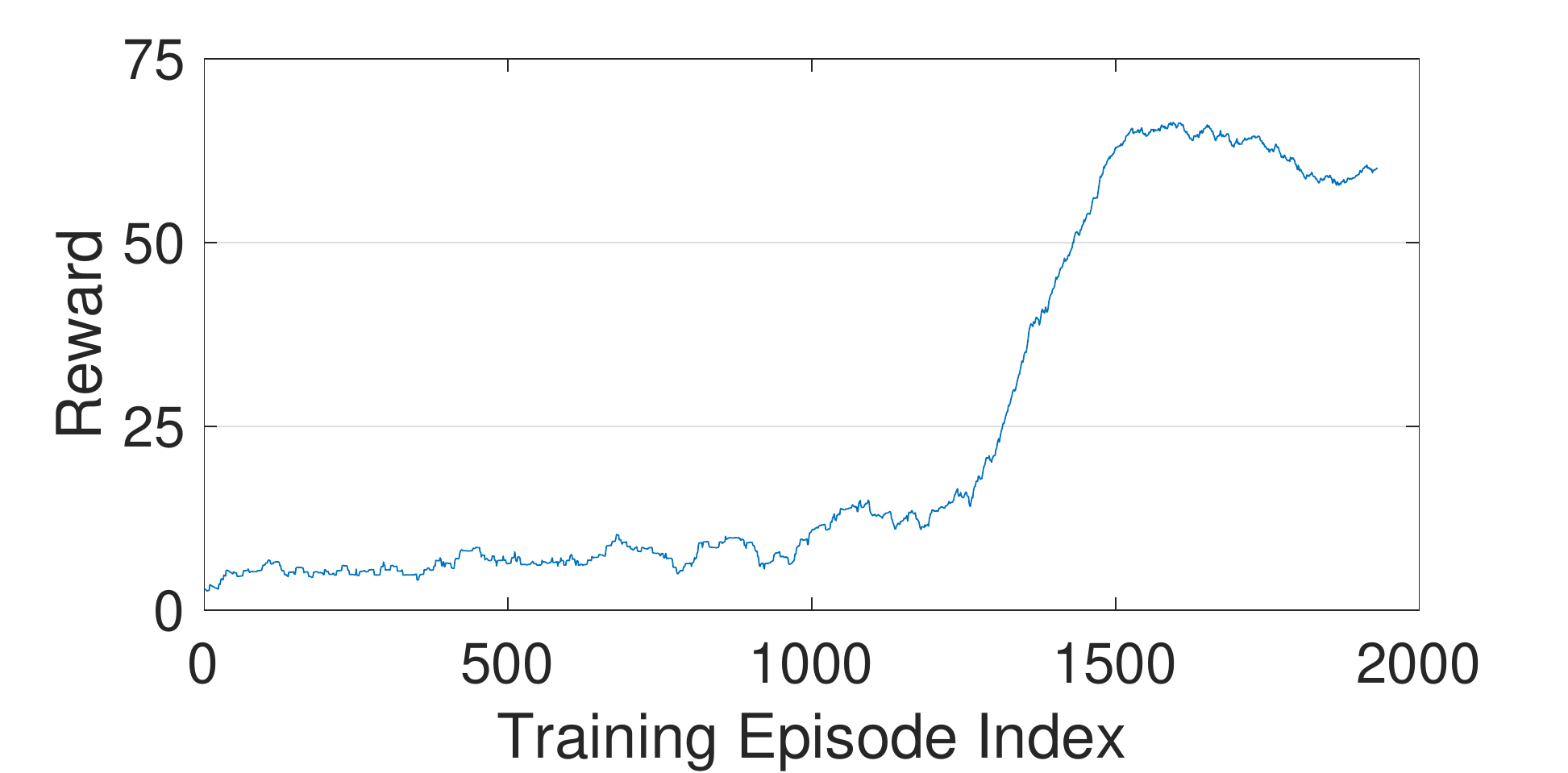}
	\caption{Reward collected during the training phase of the agent.}
	\label{fig:reward}
\end{figure}
Fig. \ref{fig:reward} shows the running average (with a window length of 100 episodes) of the reward during the training process of the agent. The figure shows that the proposed DRL approach allows to converge to a bandwidth occupancy of around 35\% (65\% of bandwidth left to other usages). During an initial exploration phase, the agent is not able to address the QoS requirement, hence the low reward. After approximately 1200 training episodes, rewards begin to grow, since the algorithm successfully learned how to satisfy the latency constraint.

The following figures are obtained by running the agent obtained at the end of the learning phase over the dataset considered for the testing phase.  

\begin{figure}[!h]
	\centering
	\includegraphics[width=0.85\linewidth]{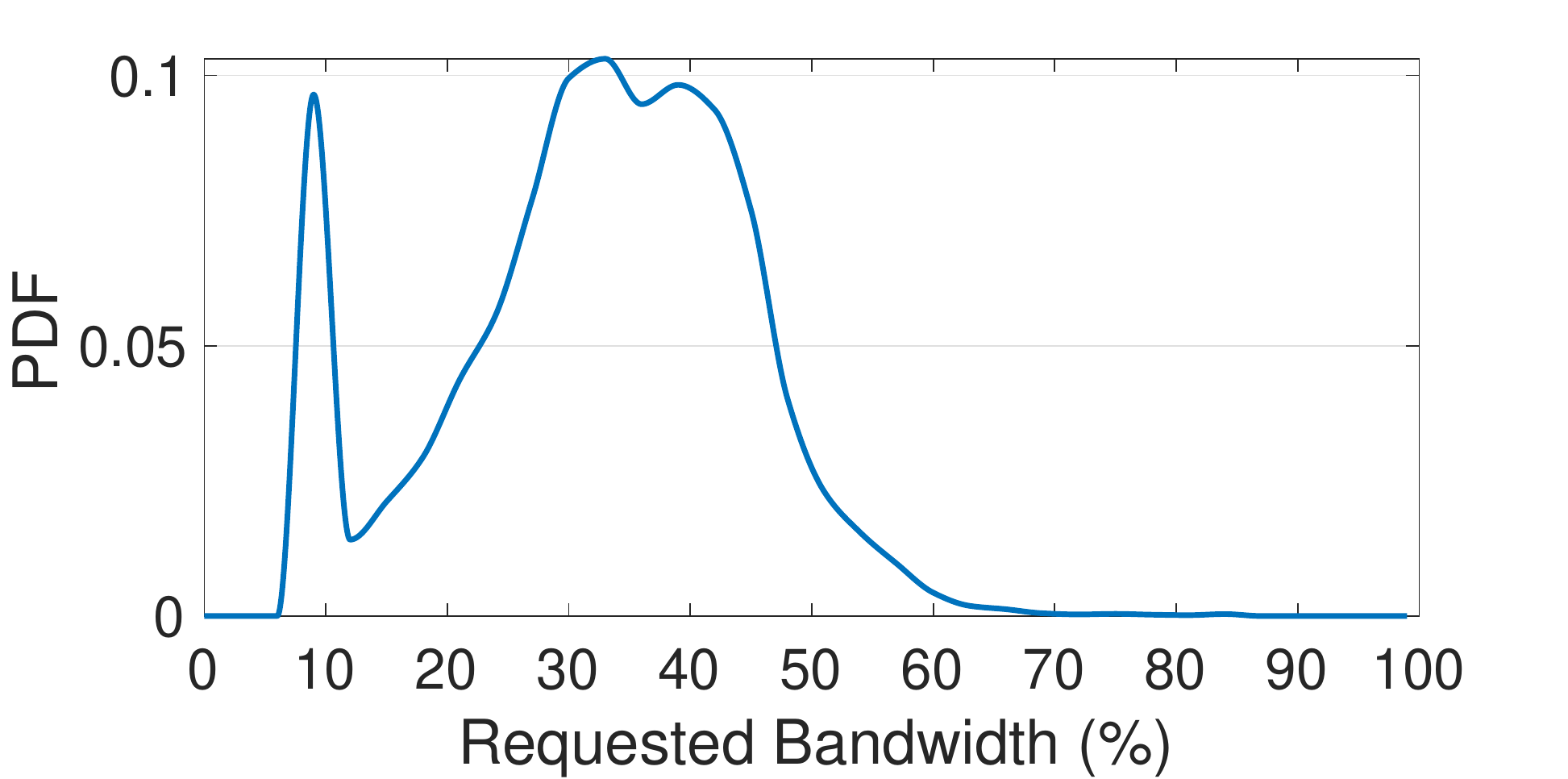}
	\caption{Probability Density Function of the actions taken by the agent.}
	\label{fig:plot2}
\end{figure}
Fig. \ref{fig:plot2} shows the probability density function of the bandwidth requested by the DRL agent of the TNT. Samples are related to 10000 independent simulations.
On the one hand, it demonstrates how the agent effectively learned to perform a variety of actions, i.e., it learned to dynamically adapt to the environment.
On the other hand, it shows that the agent usually tries to request as low bandwidth as possible, hence indicating a well-engineered reward function. 

In order to better demonstrate the importance of DRL, we compare the simulation results with the following methods:
\begin{itemize}
  \item Fixed Allocation, in which the TNT requests always the same amount of bandwidth;
  \item Heuristic strategy, characterized by a perfect prediction (i.e., ideal) of the incoming traffic; at each step the bandwidth to request is directly proportional to the incoming traffic. 
  \item Optimum allocation, in which at each step the minimum bandwidth allowing to fulfill the slice QoS requirements is determined through iterative adjustment. Clearly, this approach is unfeasible in a real system, although it can be easily simulated. 
\end{itemize}

\begin{figure}[!h]
	\centering
	\includegraphics[width=0.85\linewidth]{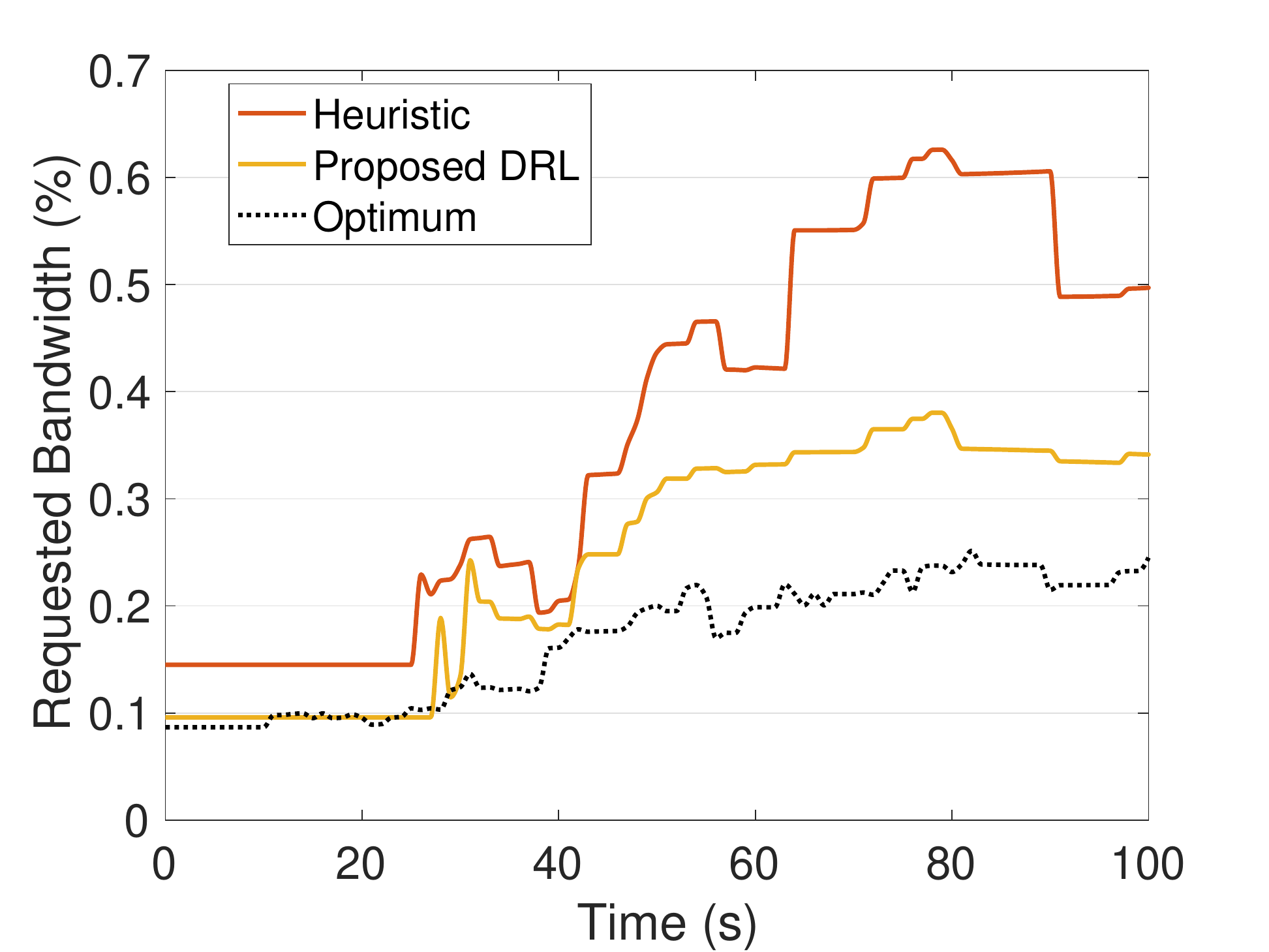}
	\caption{Bandwidth requests during a representative test episode.}
	\label{fig:plot3}
\end{figure}
Fig. \ref{fig:plot3} shows the bandwidth requested by the TNT during a representative test episode.
It clearly illustrates how the agent learned to request an amount of bandwidth close to the optimum, by taking into account only the state variables. In other words, the agent is dynamically adapting to the changes occurring in the environment.
Furthermore, it is of the utmost importance to highlight how the proposed DRL solution outperforms the heuristic approach.
In other words, even though the prediction of the incoming traffic is accurate, it is not sufficient to guarantee an optimal bandwidth request. As a matter of fact, it is necessary to take into account what actually happens in the \gls{ran} to accomplish such a decision. 
For instance, it is clear that the incoming traffic grows substantially after 60 s. However, it is reasonable to assume that general radio channel conditions improve as well, therefore it is not strictly necessary to claim more bandwidth.

\begin{figure}[!h]
	\centering
	\includegraphics[width=0.85\linewidth]{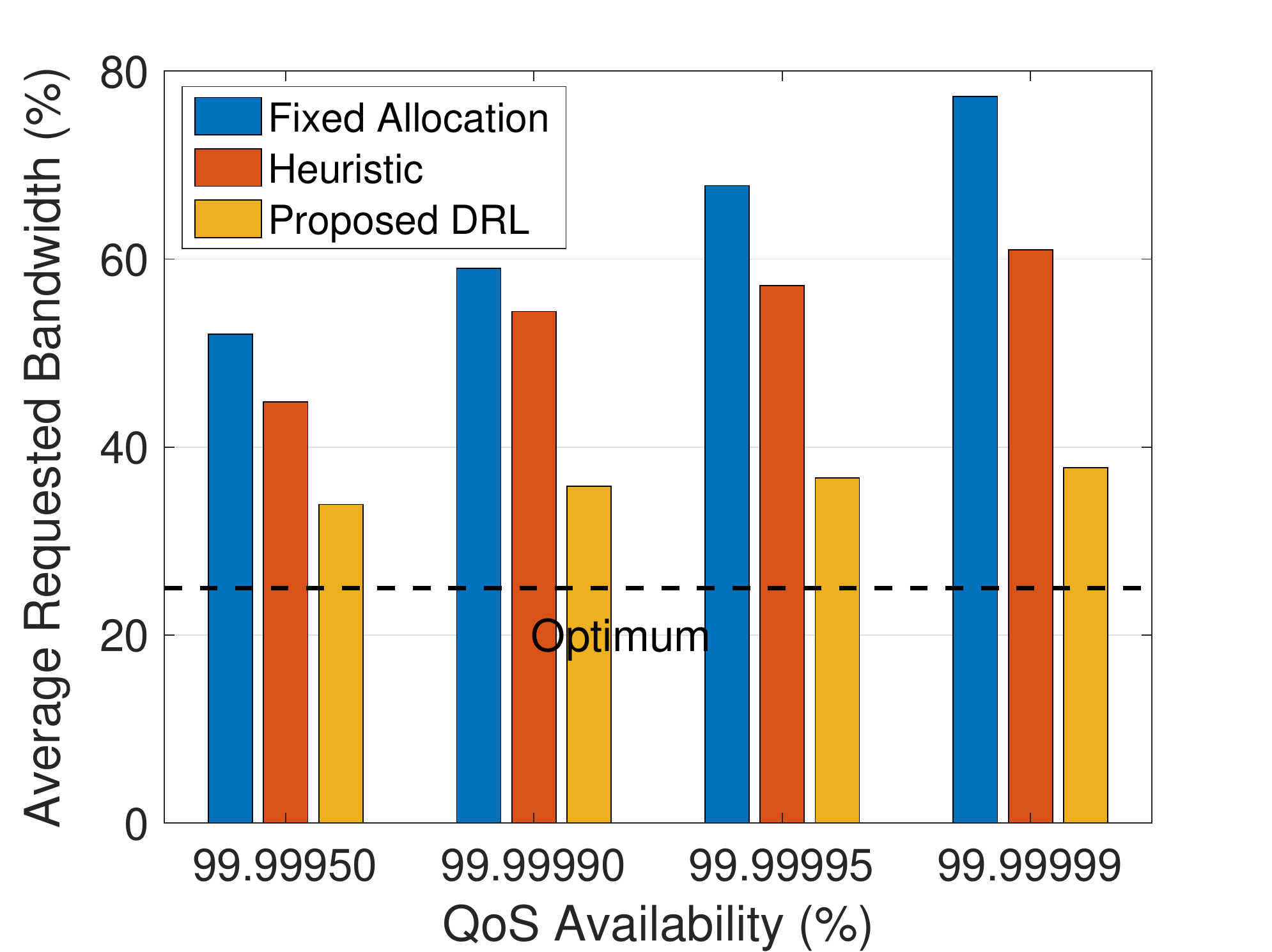}
	\caption{Average bandwidth needed to mantain a given QoS availability.}
	\label{fig:plot1}
\end{figure}
Fig. \ref{fig:plot1} shows the bandwidth requested by the TNT to ensure a certain level of QoS availability, i.e., the probability associated with the main QoS requirement being satisfied. Specifically, the actions taken by both the trained DRL agent and the heuristic are successively weighted to obtained different behaviors.
The results are then averaged over 10000 independent simulations.
The most noticeable feature is that the proposed DRL mechanism always outperforms the other strategies.
Even though requested bandwidth always grows with more stringent requirements on the QoS Availability probability, the DRL agent requests up to 50\% less bandwidth compared to the fixed allocation.
Moreover, the variation of the bandwidths requested by the TNT DRL agent are incredibly smaller, confirming how the agent learned a near-optimal allocation strategy starting from the limited information available.

\section{Conclusions and Open Challenges}
\label{Conclusions}
In this paper, we presented a novel architecture in which both the Infrastructure Provider and tenants interact for enforcing Network Slices in the next-generation \gls{ran}. Specifically, it exploits Deep Reinforcement Learning at the edge for supporting effective enforcing of \gls{ran} slicing, where tenants are encouraged to take control having an only partial understanding of the underlying RAN status.
Besides, focusing on the autonomous-driving use case, our proposal's effectiveness against baseline methodologies is investigated through computer simulation. Results confirm that the prediction of the incoming agglomerated per-slice traffic is not sufficient for an effective \gls{ran} slice enforcement strategy and resource over-provisioning is remarkably inefficient. 
Even though the proposed solution proves its success, there are still different open challenges to deal with in future works. For instance, it is important to use cutting-edge methodologies (e.g., transfer learning) for developing flexible and interoperable software agents, in order to guarantee reconfigurability and continuous deployment. Moreover, the privacy concerns of users and regulatory bodies should be also taken into account.
Finally, how to provide sufficient and powerful resources for running AI at the edge in an economically sustainable way,  is one of the fundamental challenges to be addressed to properly prepare for the advent of \gls{ei}. 
\ifCLASSOPTIONcaptionsoff
  \newpage
\fi


\begin{thebibliography}{99}


\bibitem{edge-intelligence} Z. Zhou, X. Chen, E. Li, L. Zeng, K. Luo, and J. Zhang, ``Edge intelligence: Paving the last mile of artificial intelligence with edge computing'', 2019. Proceedings of the IEEE.


\bibitem{foukas2017} X. Foukas, G. Patounas, A. Elmokashfi, and M. K. Marina, ``Network Slicing in 5G: Survey and Challenges'', IEEE Communications Magazine, vol. 55:5, pp. 94-100, May 2017.

\bibitem{martiradonna2019architecting} S. Martiradonna, A. Abrardo, M. Moretti, G. Piro, and G. Boggia, ``Architecting RAN Slicing for URLLC: Design Decisions and Open Issues'', Proc. of IEEE/ACM International Symposium on Distributed Simulation and Real Time Applications (DS-RT), Cosenza, Italy, Oct., 2019

\bibitem{elayoubi2019} S. E. Elayoubi, S. B. Jemaa, Z. Altman and A. Galindo-Serrano, ``5G RAN Slicing for Verticals: Enablers and Challenges'', in IEEE Communications Magazine, vol. 57, no. 1, pp. 28-34, January 2019, doi: 10.1109/MCOM.2018.1701319.

\bibitem{sutton} Richard S. Sutton and Andrew G. Barto. 2018. ``Reinforcement Learning: An Introduction''. A Bradford Book, Cambridge, MA, USA.

\bibitem{Zhou2016} Zhou et al., "Network Slicing as a Service: Enabling Enterprises' Own Software-Defined Cellular Networks," IEEE Commun. Mag., vol. 54, no. 7, July 2016, pp. 146-53.

\bibitem{Sallent2017} Oriol Sallent, Jordi Perez-Romero, Ramon Ferrus, and Ramon Agusti,  "On Radio Access Network Slicing From a Radio Resource Management Perspective", IEEE Wireless Communications, October 2017.

\bibitem{Konstantinos2016} Konstantinos Samdanis, Xavier Costa-Perez, and Vincenzo Sciancalepore, "From Network Sharing to Multi-Tenancy: The 5G Network Slice Broker," IEEE Communications Magazine, July 2016.

\bibitem{Capone2019} O, U. Akgul, I. Malanchini, and A. Capone, "Dynamic Resource Trading in Sliced Mobile Networks," IEEE Transactions on Network and Service Management, vol. 16, no. 1, pp. 220-233, March 2019.

\bibitem{Doro2020} Salvatore D'Oro, Francesco Restuccia, and Tommaso Melodia, ``Toward Operator-to-Waveform 5G Radio Access Network Slicing'',  IEEE Commun. Mag., April 2020.

\bibitem{Nikravesh2016} A. Y. Nikravesh et al., ``An Experimental Investigation of Mobile Network Traffic Prediction Accuracy'', Services Transactions on Big Data, vol. 3, no. 1, pp. 1-16, Jan. 2016.

\bibitem{Bega2020} Dario Bega, Marco Gramaglia, Marco Fiore, Albert Banchs, and Xavier Costa-Perez, ``DeepCog: Optimizing Resource Provisioning in Network Slicing with AI-based Capacity Forecasting'', JSAC February 2020.

\bibitem{Fu2016} F. Xu et al., ?Big Data Driven Mobile Traffic Understanding and Forecasting: A Time Series Approach,? IEEE Transactions on Services Computing, vol. 9, no. 5, pp. 796?805, Sep. 2016.

\bibitem{Zhang2017} M. Zhang et al., ?Understanding Urban Dynamics From Massive Mobile Traffic Data,? IEEE Transactions on Big Data, vol. 5, no. 2, pp. 266? 278, Nov. 2017.

\bibitem{Li2014} R. Li et al., ?The prediction analysis of cellular radio access network traffic: From entropy theory to networking practice,? IEEE Communications Magazine, vol. 52, no. 6, pp. 234?240, Jun. 2014.

\bibitem{Zhang2018} C. Zhang et al., ?Long-Term Mobile Traffic Forecasting Using Deep Spatio-Temporal Neural Networks,? in Proc. of ACM MobiHoc, Los Angeles, CA, USA, Jun. 2018, pp. 231?240.

\bibitem{Gutterman2019} C. Gutterman et al., ?RAN resource usage prediction for a 5G slice broker,? in Proc. of ACM Mobihoc, Catania, Italy, Jul. 2019.

\bibitem{Ye2019} Hao Ye , Geoffrey Ye Li, and Biing-Hwang Fred Juang, "Deep Reinforcement Learning Based Resource Allocation for V2V Communications," IEEE Transactions in Vehicular Technology, Vol. 68., No. 4, April 2019.

\bibitem{Faris2020} Faris B. Mismar , Brian L. Evans , and Ahmed Alkhateeb , ``Deep Reinforcement Learning for 5G Networks: Joint Beamforming, Power Control, and Interference Coordination'', IEEE Transactions on Communications, VOL. 68, NO. 3, March 2020.

\bibitem{5gaa} 5GAA Automotive Association, ``C-V2X Use Cases: Methodology, Examples and Service Level Requirements'', 2019.

\bibitem{itu} International Telecommunication Union Radiocommunication sector (ITU-R), ``Guidelines for evaluation of radio interface technologies for IMT-2020'', Report M.2083, 11 2017, [online] Available: https://www.itu.int/pub/R-REP-M.2412.

\bibitem{dataset} M. Piorkowski, N. Sarafijanovic‑Djukic, and M. Grossglauser, ``CRAWDAD dataset epfl/mobility (v. 2009‑02‑24)'', downloaded from https://crawdad.org/epfl/mobility/20090224, https://doi.org/10.15783/C7J010, Feb 2009.














\end{thebibliography}
\end{document}